\documentclass[aps,prb,superscriptaddress,twocolumn,amsmath,amssymb,floatfix]{revtex4}
\usepackage{tabularx}
\usepackage{bm}
\usepackage{euscript}
\usepackage{epsfig,psfrag,subfigure}
\usepackage{graphicx}
\usepackage{color}
\usepackage{amsfonts}
\usepackage{exscale}
\usepackage{amsbsy}
\usepackage{hyperref}
\def\prb{Phys. Rev. B}

\def\be{\begin{equation}}
\def\ee{\end{equation}}
\def\ba{\begin{eqnarray}}
\def\ea{\end{eqnarray}}

\begin{document}

\title{The Optimal Inhomogeneity for Superconductivity: Finite Size Studies}
\author{Wei-Feng Tsai}
\affiliation{Department of Physics and Astronomy, University of California, Los Angeles, CA 90095}
\affiliation{Department of Physics, Stanford University, Stanford, CA 94305}
\author{Hong Yao}
\affiliation{Department of Physics, Stanford University, Stanford, CA 94305}
\author{Andreas L\"auchli}
\affiliation{Institut Romand de Recherche Num\'erique en Physique des Mat\'eriaux (IRRMA), CH-1015 Lausanne, Switzerland}
\author{Steven A. Kivelson}
\affiliation{Department of Physics, Stanford University, Stanford, CA 94305}

\date{\today}

\begin{abstract}
We report the results of exact diagonalization studies of Hubbard models on a $4\times 4$ square lattice with periodic boundary conditions and various degrees and patterns of inhomogeneity, which are represented by inequivalent hopping integrals $t$ and $t^{\prime}$. 
We focus primarily on two patterns, the checkerboard and the striped cases, for a large range of values of the on-site repulsion $U$ and doped hole concentration, $x$. 
We present evidence that superconductivity is strongest for $U$ of order the bandwidth, and intermediate inhomogeneity, $0 < t^\prime <  t$.    
The maximum value of the ``pair-binding energy'' we have found with purely repulsive interactions is $\Delta_{pb} = 0.32t$ for the checkerboard Hubbard model with $U=8t$ and $t^\prime = 0.5t$.  Moreover, for near optimal values, our results are insensitive to changes in boundary conditions, suggesting that the correlation length is sufficiently short that finite size effects are already unimportant.
\end{abstract}

\maketitle

The relatively large energy scales and short coherence lengths involved in high temperature superconductivity (HTC) imply that theories of the ``mechanism'' must involve different considerations than the conventional BCS theory of low temperature superconductivity (LTC).  The theory of LTC can be treated in the context of Fermi liquid theory, in which the strong effects of electron-electron repulsions are resolved at high energy, so that pairing is a Fermi surface instability triggered by weak, retarded, induced attractive interactions.  The theory of HTC must treat the strong local repulsions between electrons directly, as Fermi liquid theory is certainly not valid at short distances and high energies.  Conversely, since the physics of HTC is relatively local, numerical studies of finite size systems can, plausibly, resolve questions concerning the mechanism so long as the system sizes are large compared to the coherence length, a condition that could not remotely be envisaged for LTC.

\begin{figure}[t]
\includegraphics[width=6.6cm]{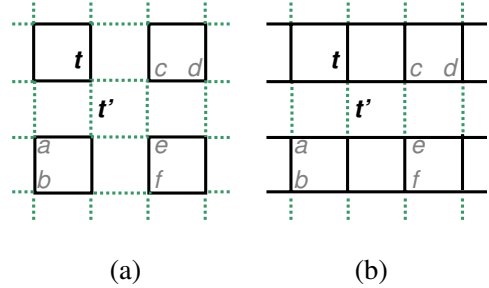}
\caption{{\footnotesize (Color online) Schematic representation of the 
inhomogeneous Hubbard model: 
(a) the checkerboard version and (b) the striped version.
The hopping matrix elements are $t$ on the solid ``strong'' bonds and $t^\prime \leq t$ on the dotted ``weak'' bonds. The three labeled bonds, $\overline{ab}$, $\overline{cd}$, and $\overline{ef}$
are the focus of our study of pairing correlations.}}
\label{fig:lattice}
\end{figure}

In the present paper, we report exact diagonalization studies of the low energy states of the Hubbard model on a square lattice (Fig.~\ref{fig:lattice}), with  hopping matrix elements, $t_{ij}$, between pairs of sites, $i$ and $j$, and with a short-range repulsive interaction, $U_j$, between two electrons on the same site.  We propose to address the following sharply posed question, related to the physics of the mechanism:  What form of the Hamiltonian ({\it i.e.} values of $\{t_{ij}\}$ and $\{U_i\}$)  maximizes $T_c$ subject to the constraints that 1) $t_{ij}$ is short-ranged and bounded, {\it i.e.} $|t_{ij}| \leq t$ for all $ij$, and that 2) the interactions are repulsive, $U_i \geq 0$ for all~$i$?  Constraint 2 represents a theoretical prejudice that HTC derives directly from the strong repulsive interactions between electrons, and could be relaxed in future studies.  The requirement that $t_{ij}$ is bounded avoids the trivial answer that, for any Hamiltonian that supports superconductivity,  $T_c$ can be doubled by simply doubling the Hamiltonian.  

The largest systems we can readily study are 4 $\times$ 4.  For such small systems, there is no direct way to extract a $T_c$.  We have thus introduced other benchmarks of the strength of the superconductivity that can be readily computed on finite size systems, especially the ``pair binding energy,'' defined in Eq.~(\ref{fig:pbenergy}), and the strength of the pair-field correlations defined in Eqs.~(\ref{Dij})-(\ref{avDelta}). 

The full optimization problem we have proposed would be prohibitively time consuming.  We have therefore concentrated on a restricted, highly symmetric subset of all possible Hamiltonians, with particular focus on the symmetric checkerboard and stripe patterns shown in Figs.~\ref{fig:lattice}(a)  and \ref{fig:lattice}(b). 
We take the hopping matrix elements to be $t$ on the solid (``strong'') bonds in the figure, and $t^\prime \leq t$ on the dotted (``weak'') bonds.
 For $t^\prime = t$, the system is the homogeneous Hubbard model, while for $t^\prime=0$, the system consists of disconnected Hubbard squares or ladders.  Thus, as we vary $t^\prime$ from 0 to $t$, we vary the ``degree of inhomogeneity''.  
 
\begin{figure}[tp]
\subfigure[$\Delta_{pb}(1/16)$, checkerboard lattice]{\includegraphics[width=7.6cm]{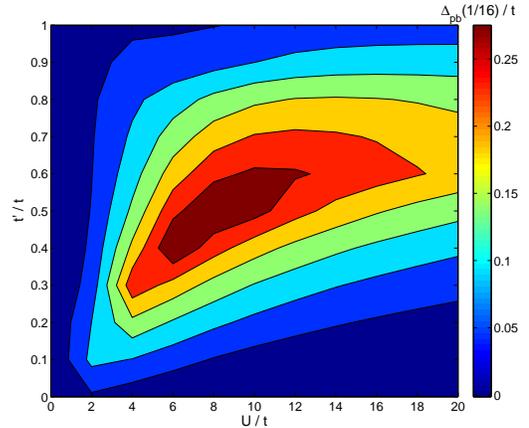}} 
\subfigure[$\Delta_{pb}(3/16)$, checkerboard lattice]{\includegraphics[width=7.6cm]{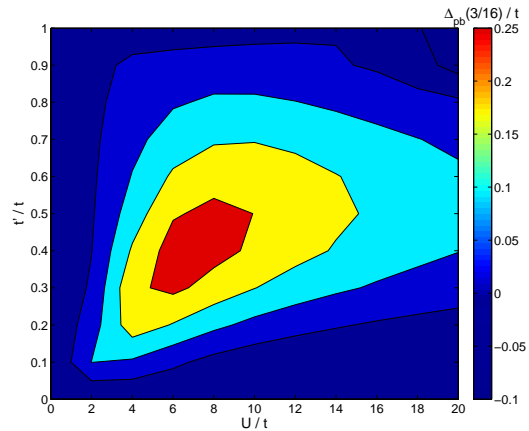}}
\subfigure[$\Delta_{pb}(1/16)$, striped lattice]{\includegraphics[width=7.6cm]{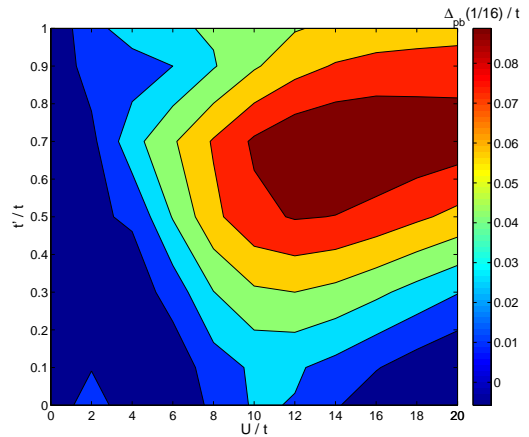}}  
\caption{{\footnotesize (Color online) Contour plots of the pair-binding energy as a function of $U$ and $t^{\prime}$ on two types of lattices with periodic boundary conditions.   Note change of scale in c.  
}}
\label{fig:contour_pbe}
\end{figure}

 For instance, as shown in the contour plot in Fig.~\ref{fig:contour_pbe}, the pair-binding energy of the checkerboard lattice with periodic boundary conditions (PBC) is largest for $U=8t$ and $t^\prime = 0.5 t$.  
The concentration, $x$, of ``doped holes'' per site ({\it i.e.} the deviation from one electron per site) can only take 
 discrete values;  among those, the optimal pair-binding is largest when $x=1/16$ (as in the contour plot in Fig.~\ref{fig:contour_pbe}(a)), somewhat smaller when $x=3/16$ (Fig.~\ref{fig:contour_pbe}(b)), and is  always negative (pair-repulsion) for $x=5/16$.  Indeed, of all forms of the Hamiltonian we have explored to date, the checkerboard Hubbard model with these parameters has the largest pair-binding energy we have found.  Moreover, by changing boundary conditions to twisted PBC, shown as the solid (red) triangles in Fig.~\ref{fig:pbenergy}(a), we see that at fixed $U=8t$, as a function of $t^\prime$ in the range $ 0.8\gtrsim  t^\prime/t \geq 0 $, the results are 
sensitive to change of boundary conditions only at the 20\% level.          

 We therefore infer that,  {\it  the checkerboard Hubbard model in the thermodynamic limit, has a 
  maximum value of the pair-binding $\Delta_{pb} \approx  t/3$ for  $U\approx 8t$, $t^\prime \approx t/2$, and $x\approx 1/16$.}  From analysis of the ground-state symmetry and of the pair-field correlations, we can identify the dominant superconducting correlations on this system as d-wave ($d_{x^2-y^2}$).   

\section{The Hamiltonian}
The inhomogeneous Hubbard model we have studied is described by the Hamiltonian 
\be \label{eq:ihm}
H=-\sum_{\left\langle i,j\right\rangle,\sigma}t_{ij}
\left(c^{\dagger}_{i,\sigma}c_{j,\sigma}+ h.c.\right)+ \sum_{i}U_{i}n_{i,\uparrow}n_{i,\downarrow},
\ee
where 
$\left\langle i,j\right\rangle$ indicates nearest-neighbor sites, 
 $c^{\dagger}_{i,\sigma}$ creates an electron on site $i$ with spin polarization $\sigma=\pm 1$, 
 and $n_{i,\sigma}=c^\dagger_{i,\sigma}c_{i,\sigma}$. The usual (homogeneous) limit of this model is obtained by taking $t_{ij}=t$ and $U_{i}=U$. 

Although we have studied a wider variety of patterns, to be concrete we will primarily focus on two 
 inhomogeneous patterns of the hopping amplitudes: 
  the checkerboard lattice and the striped lattice 
   shown in Fig.~\ref{fig:lattice}. 
   Unless otherwise stated, our discussion will focus on the case in which PBCs are applied in 
  both the $x$ and $y$ directions. 
  However, finite size effects will be estimated by comparing 
  these results to those  with 
  ``twisted'' periodic boundary conditions.  Specifically, for a 4 $\times$ 4 system,  PBC means identifying the sites  $(n+4,m)\equiv (n,m)$ and $(n,m+4)\equiv (n,m)$, while  twisted PBC in the y direction means identifying  $(n+4,m)\equiv (n,m)$ and $(n,m+4)\equiv (n+2,m)$.  
  We have also obtained results with open boundary conditions, but because of the large surface to volume ratio of the small systems studied, the proper interpretation of these results is unclear, and so we do not report them, here.

\section{Results}
 The exact diagonalization is performed using the Lanczos method~\cite{Cullum85}, which has been employed by other people to extract some ground state properties of a $4\times 4$ (homogeneous) Hubbard model~\cite{Dagotto92,Fano90,Fano89}. By taking advantage of conservation of the $z$ component of the spin and the $C_{4v}$ symmetry of the checkerboard and $D_{2h}$ symmetry of the striped model, we successfully reduce the dimension of the Hilbert space from around $10^{8}$ to $10^{7}$ states and the ground state energies we found for $t^{\prime}/t\rightarrow 1$ agree with those of G.~Fano {\it et al.}~\cite{Fano90}. A complete group theoretical analysis which facilitates efficient implementation of the Lanczos algorithm can be found in Ref.~\onlinecite{Fano89}.    

We present representative results on the {\it pair-binding energy} and  the {\it pair-field correlations}. 
We have much more extensive tables of results for various values of $U/t$, $t^\prime/t$ and $x$, and for various choices of boundary conditions.  
 These additional results are available from the authors upon request.

\subsection{The Pair-Binding Energy}
To better understand the pairing phenomena arising from repulsive interactions, we define the 
 pair-binding energy:
\be \label{eq:pbenergy}
\Delta_{pb}(x)=2E_0{(M)} - [E_0{(M+1)}+E_0{(M-1)}],
\ee
where $N=16$ is the number of sites in the system, $E_0{(M)}$ is the ground state energy with $N-M$ electrons ({\it i.e.} $M$ holes doped into a ``neutral'' 
half-filled lattice), and $x\equiv M/N$ is the ``concentration of doped holes.''  
We will focus on the case in which $M$ is odd.  Thus, a
 positive pair-binding energy means that, given two isolated clusters with a mean doped hole density $x$, it is energetically preferable to ``pair'' the doped holes so that one cluster has $M+1$ and the other cluster has $M-1$ doped holes.  
 
For a superconducting system, $\Delta_{pb} \to 2\Delta_{min}$ in the limit $N\to \infty$, where $\Delta_{min}$ is the minimum  value of the superconducting gap.  While gaps can occur for other reasons ({\it e.g.}  CDW formation), a superconducting state, as far as we know, is the unique state that generically produces a gap for a non-zero range of $x$ in more than 1D.  In a superconducting state with gapless (nodal) excitations, $\Delta_{pb}$ vanishes as $N\to \infty$, but only relatively slowly, in proportion to $\Delta_{0}N^{-1/2}$ in 2D, so $\Delta_{pb}$ computed on finite size systems remains a good diagnostic of superconductivity.
Under generic circumstances in non-superconducting systems, the repulsive interactions between quasiparticles implies that $\Delta_{pb}$ is negative and 
$\Delta_{pb} \sim -N^{-1}$ as $N\to\infty$.

A positive pair-binding energy on a small system could also indicate a tendency for phase separation.  Unambiguously distinguishing gap formation from phase separation can only be done by appropriate finite size scaling\cite{PhysicaC}, which is beyond the reach of the present calculations.  However, a gross test for phase separation is possible by testing whether further agglomeration of doped holes is favored.  Specifically, for $M=Nx$ even,
we define 
\begin{equation}
\kappa_{N}=[E_0(M+2)+E_0(M-2)-2E_0(M)]/2 .
\end{equation}
This is a crude approximation of the compressibility\cite{PhysicaC}, which is  negative in a system with a sufficiently strong tendency  to phase separation.   For the interesting range of $t^{\prime}$ and $U$, we 
 never find a negative value of $\kappa_{N}$ and hence our system is unlikely to be phase separated.  

\begin{figure}[tp]
\subfigure[]{\includegraphics[width=7.8cm]{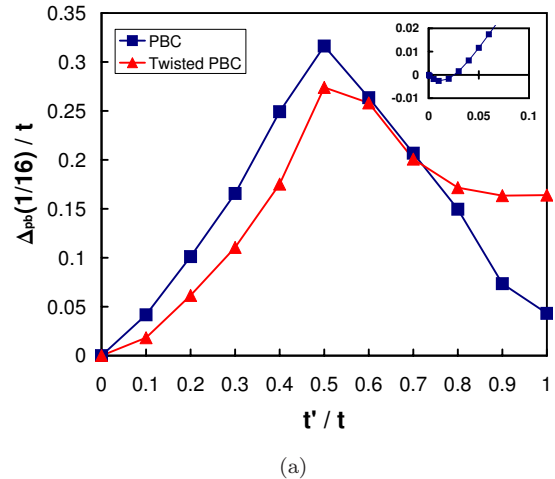}} 
\subfigure[]{\includegraphics[width=7.8cm]{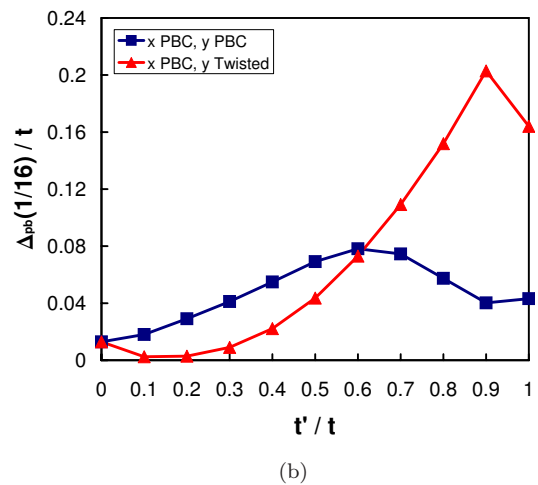}} 
\caption{{\footnotesize (Color online) The pair-binding energy, $\Delta_{pb}(1/16)$, as a function of $t^{\prime}$ at $U=8t$ on a (a)~checkerboard lattice and a (b)~striped lattice with various boundary conditions. Squares represent data with PBC in both directions; triangles represent data with PBC in the $x$ direction and ``twisted'' PBC in the $y$ direction.  [Inset: A closer look at $\Delta_{pb}(1/16)$ as $t^\prime$ goes to zero on the checkerboard lattice. Notice that $\Delta_{pb}(1/16)$ becomes negative when $t^\prime \lesssim 0.025t$.]}}
\label{fig:pbenergy}
\end{figure} 

We have computed the pair-binding energy $\Delta_{pb}$ 
as a function of $t^{\prime}$, 
$U$, and $x$
on both 
the checkerboard and striped lattices with PBCs and twisted PBCs.  
Contour plots of $\Delta_{pb}(x=1/16)$ and $\Delta_{pb}(x=3/16)$ for the checkerboard lattice are shown in respectively, in Figs.~\ref{fig:contour_pbe} (a) and (b), respectively. The global maximum can clearly be seen for $U=8t$ and $t^\prime=0.5t$.

\begin{figure}[t]
\includegraphics[width=7.2cm]{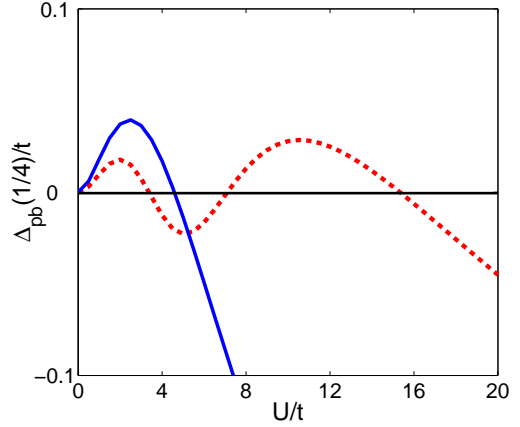} 
\caption{{\footnotesize (Color online) The pair-binding energy, $\Delta_{pb}(1/4)$, as a function of $U$ for an isolated $2\times 2$ plaquette (solid curve) and a $4\times 2$ ladder (dashed curve). 
Note that in the $U=\infty$ limit one can show $\Delta_{pb}(1/4)\le 0$ for both clusters\cite{sudip91}.}}
\label{fig:cell_pbe}
\end{figure}

$\Delta_{pb}(x=1/16)$ is shown at fixed $U=8t$ as a function of $t^\prime$ in Fig.~\ref{fig:pbenergy} with both PBC (squares) and twisted PBCs (triangles).  A remarkable degree of insensitivity to boundary conditions is apparent for $t^\prime/t \leq 0.8$. On the coarse scale of $t^\prime$ in the main figure, it appears that $\Delta_{pb}$ is positive at small $t^\prime/t$ for all $U$.  However, this hides a subtlety at small $t^\prime/t$, as shown in the inset to the figure in which the regime of small $t^\prime$ is shown on an expanded scale. Specifically, as shown in Fig.~\ref{fig:cell_pbe}, the pair-binding energy on an isolated square changes from positive to negative at $U=U_c\approx 4.6t$.  Moreover, as discussed in Ref.~\onlinecite{Tsai06}, it follows that the pair binding is positive with a non-zero limit as $t^\prime/t \to 0$ for $0 < U < U_c$, while $\Delta_{pb} \sim -{\cal O}([t^\prime]^2/t)$ is negative and vanishes as $t^\prime/t \to 0$ for $U > U_c$.

\begin{figure}[t]
\includegraphics[width=7.8cm]{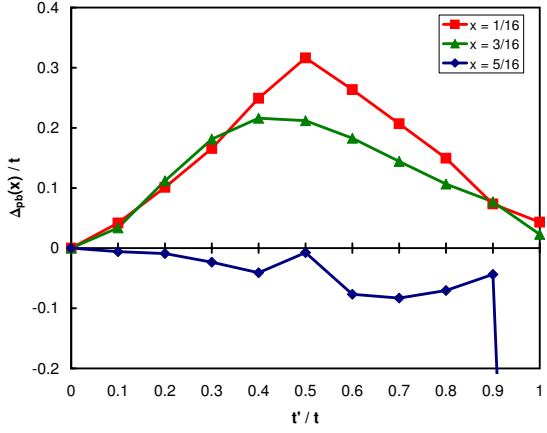} 
\caption{{\footnotesize (Color online) The doping dependence of the pair-binding energy, $\Delta_{pb}(x)$, as a function of $t^\prime$ on the checkerboard lattice.}}
\label{fig:doping_pbe}
\end{figure}

In Fig.~\ref{fig:doping_pbe}, $\Delta_{pb}(x)$ for the checkerboard model is shown at fixed $U/t=8$ as a function of $t^\prime/t$ for different values of $x =1/16$, 3/16, and 5/16.  

A contour plot of $\Delta_{pb}(x=1/16)$  for the striped  lattice is shown in  in Fig.~\ref{fig:contour_pbe} (c), and $\Delta_{pb}(x=1/16)$ for fixed $U=8t$ is shown as a function of $t^\prime/t$ for PBC and twisted PBC 
 in Fig.~\ref{fig:pbenergy} (b).
Here, the results are apparently more sensitive to boundary conditions, so inferences concerning the thermodynamic limit are more difficult to reach.  However, here, too, a global maximum of $\Delta_{pb}$ is reached for $U=14t$ and $t^\prime = 0.7t$.  In the limit of vanishing $t^\prime$, as follows from the results for an isolated ladder, Fig.~\ref{fig:cell_pbe} (dashed curve), $\Delta_{pb}$ is positive and non-vanishing for $0 < U < U_{c1}\approx 3.5t$, negative and order $(t^\prime)^2$ for $U_{c1} < U < U_{c2}\approx 7t$, again positive and non-vanishing for $U_{c2} < U < U_{c3}\approx 15t$, and finally negative and order $(t^\prime)^2$ for $U > U_{c3}$.

\subsection{Pair-field correlations}

We have also studied the equal-time pair-field pair-field correlation function defined as
\be
D(\overline{ij},\overline{kl})=\langle \Delta^{\dagger}_{ij}\Delta_{kl}\rangle,
\label{Dij}
\ee
where the pair-field is
\be
\Delta^{\dagger}_{ij}=\frac{1}{\sqrt{2}}(c^{\dagger}_{i\uparrow}
c^{\dagger}_{j\downarrow}+c^{\dagger}_{j\uparrow}c^{\dagger}_{i\downarrow}).
\label{Deltaij}
\ee
$\overline{ij}$ 
represents the bond between a pair of nearest-neighbor sites, $i,j$, on the lattice.  We focus on this correlation function for the largest possible separations, given the small system size:  between a pair of parallel strong bonds separated by a distance 2, {\it i.e.} bonds $\overline{ab}$ and $\overline{ef}$ in Fig.~\ref{fig:lattice}, and a pair of perpendicular strong bonds separated by a distance $3\sqrt{2}/2$,  {\it i.e.} bonds $\overline{ab}$ and $\overline{cd}$ in Fig.~\ref{fig:lattice}.
%
We have computed these pairing correlations for both the checkerboard and striped models with 
$M=0$, 2, and 4, {\it i.e.} for doped hole concentration $x=0$, 1/8, and 1/4, and for a range of $U/t$ and $t^\prime/t$. 

\begin{figure}[t]
\subfigure[]{\includegraphics[width=7.8cm]{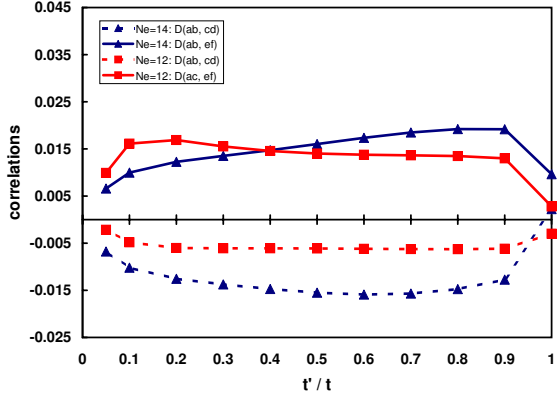}}
\subfigure[]{\includegraphics[width=7.8cm]{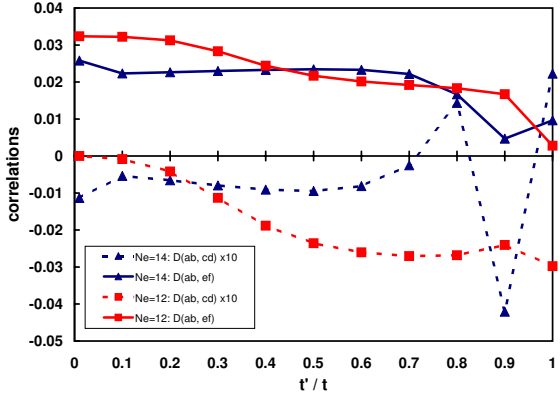}}
\caption{{\footnotesize (Color online) Pair-field pair-field correlation functions at $U=8t$ on the (a) checkerboard and (b) striped lattices. Solid curves represent $D(\overline{ab},\overline{ef})$ and dashed represent $D(\overline{ab},\overline{cd})$; triangles are data points for $M=2$ doped holes, and squares are data points for $M=4$. 
Note that in the striped case $D(\overline{ab},\overline{cd})$ have been multiplied by ten for comparison.}}
\label{fig:PairCorr}
\end{figure}

In Fig.~\ref{fig:PairCorr}(a), we show the pair correlation function for the checkerboard lattice at fixed $U/t=8$ as a function of $t^\prime/t$ for $x=1/8$ (triangles) and 1/4 (squares). $D(\overline{ab},\overline{cd})$ is represented by the dashed lines in the figure, and $D(\overline{ab},\overline{ef})$ by the solid lines.
 The same quantities are shown for the striped lattice in Fig.~\ref{fig:PairCorr}(b).  Qualitatively similar results for both lattices have been obtained for $x=0$, but the mangnitude of $D$ is an order of magnitude smaller than for $x >0$, consistent with the expected Mott insulating character of the undoped system.  The positive sign of $D(\overline{ab},\overline{ef})$ and the negative sign of $D(\overline{ab},\overline{cd})$ are consistent with the d-wave character of the pairing;  in the thermodynamic, at large spatial separations, the sign of $D$ is determined entirely by the symmetry of the order parameter.  
 
 For the checkerboard model, it is apparent that the magnitude of the pairing correlations are relatively weak both in the limit of strong inhomogeneity ($t^\prime/t \ll 1$) and of vanishing inhomogeneity ($t^\prime/t \approx 1$).  However,
  the strength of these correlations is only weakly dependent on $t^\prime/t$ for a broad range of intermediate values.  Moreover, in this intermediate regime, the strength of the pairing is quite similar for $x=1/8$ and $x=1/4$.  
  
\begin{figure}[t]
\includegraphics[width=7.8cm]{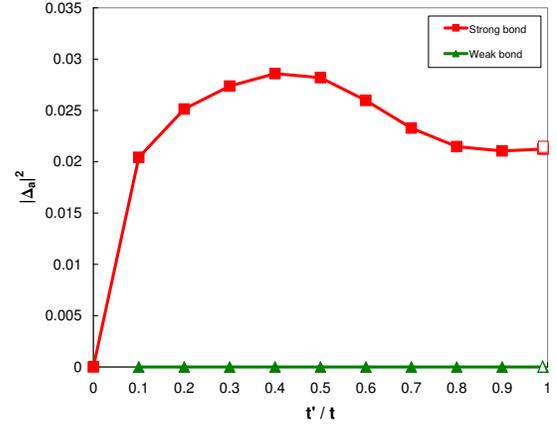} 
\caption{{\footnotesize (Color online) The square of the 
expectation value of the pair annihilation operator, $|\Delta_{a}|^2$, as a function of $t^\prime$ on the checkerboard lattice. 
The subscript $a$ refers to the contribution from the strong and weak bonds, respectively. The open symbols indicate the data obtained at $t^{\prime}=0.99t$. As in Eq.~(\ref{SMA}), $1/\epsilon$ times this quantity is the single-mode-approximation to the superconducting susceptibility.}}
\label{fig:cc_sq}
\end{figure}

  It is more difficult to make a clear qualitative statement about the behavior of the striped model.
%
%
%
One seemingly puzzling feature of the striped results deserves comment.  In the 
 limit $t^{\prime}\to 0$, where the two ladders decouple, 
 one might expect the correlation of the inter-ladder pair fields [$D(\overline{ab},\overline{cd})$] to vanish. 
 However, since for $U=8$, the pair binding 
 energy on a $4\times 2$ ladder (cube) is positive, as shown in Fig.~\ref{fig:cell_pbe} (dashed curve), for $t^{\prime}=0$ and 
 $M=2$, there are two degenerate ground states, with the hole-pair on one or the other disconnected ladder.  
 Thus, even as $t^\prime$ tends to 0, 
 the ground-state wave function is a coherent superposition of these two states, and hence has substantial pair-field correlations.  
Were one to study the finite temperature properties of this system, the coherence would be lost above a relatively small temperature $T_{coh} \sim t^\prime$.
This illustrates the dangers of uncritically accepting evidence of strong superconducting correlations from the pair-field correlation function. 


 We have also studied the expectation value of the pair-field operator between ground-states with $M=Nx$ and $M-2$ doped holes:
\be
\langle \Delta_{ij} \rangle \equiv \langle M;0 |   \Delta_{ij} | M-2;0 \rangle
\label{avDelta}
\ee
where $| M;0 \rangle$ is the ground state with $M$ doped holes.  This is not a gauge invariant quantity, in that there is an arbitrary choice of an overall phase;  by choosing the wave functions real, this is reduced to an overall sign ambiguity.  However, the internal symmetry of the pairing state is manifest in this quantity.  

For the checkerboard case,  for all $U$ and $t^\prime< t$, the ground state symmetry as a function of $M$ alternates, $A_{1}$ for $M=0$, $B_{1}$ for $M=2$, $A_{1}$ for $M=4$.  Thus, the pair-field operator that connects any two of these states must have precisely $B_{1}$ ({\it i.e.} $d_{x^2-y^2}$) symmetry\cite{trugman}.  In Fig.~\ref{fig:cc_sq}, for fixed $U=8t$,  we plot, as a function of $t^\prime/t$, the gauge-invariant quantities, $|\Delta_{s}|^2$ (squares) and $|\Delta_w|^2$ (triangles), where $\Delta_s$ and $\Delta_w$ are, respectively, the expectation value of $\Delta_{ij}$ for strong bonds $ij$ (within a square) 
and weak bonds (connecting two squares) [see Eq.~(8)]. 
 Again, there is a clear indication that superconducting correlations are strongest for $t^\prime \approx t/2$.

Two peculiarities are worth mentioning here: 
Firstly, it is puzzling 
that the difference in the pair amplitudes on the weak and strong bonds is so large, $|\Delta_w|/|\Delta_s|\sim 10^{-7}$, as clearly shown in Fig.~\ref{fig:cc_sq}.  This is a qualitative point that warrants further study.
Secondly, even in the limit $t^{\prime}\rightarrow t$, $|\Delta_w|\neq|\Delta_s|$.
This is a consequence of the fact the ground-state is three-fold 
degenerate\cite{Fano89} when $t^\prime=t$ and $M=2$;   
this degeneracy is lifted whenever $t^{\prime}\neq t$.

\subsection{Superconducting Susceptibility}
The $T=0$ superconducting susceptibility is expressible as
\be
\chi_a(\mu) = \sum_\alpha \frac {\left| \langle M ;0 |   \Delta^{(a)} | M-2;\alpha \rangle \right |^2}{E_\alpha(M-2) - E_0(M) -2\mu}
\ee
where $E_{\alpha}(M)$ is the energy of the $\alpha$th excited state with $M$ doped holes, and where we take 
\be
\Delta^{(a)} \equiv \frac 1 {Z_a} \sum_{\overline{ij}}f^{(a)}_{ij}\Delta_{ij}\ \  ; \ \ \ Z_a=\sum_{\overline{ij}} |f^{(a)}_{ij}|^2.
\ee
Here, the chemical potential is appropriate to the case in which the ground-state has $M$ doped holes, $\it{i.e.}$ $[E_0(M-2) - E_0(M)] > 2\mu > [E_0(M) - E_0(M+2)]$. Because of the d-wave symmetry, we always take $f_{ij}$ to be positive on vertical bonds and negative on horizontal bonds.  We consider three possible susceptibilities:  $\chi_s$ in which 
where, $|f_{ij}|=1$ on the strong bonds and 0 on the weak, $\chi_w$ in which $|f_{ij}|=1$ on the weak bonds and 0 on the strong, and $\chi_T$ in which $|f_{ij}|=1$ on all nearest-neighbor bonds.  For simplicity, we have defined the susceptibility with respect to adding two-holes.  Another (more conventional) definition includes, as well, terms which remove two holes.  

\begin{figure}[t]
\includegraphics[width=7.8cm]{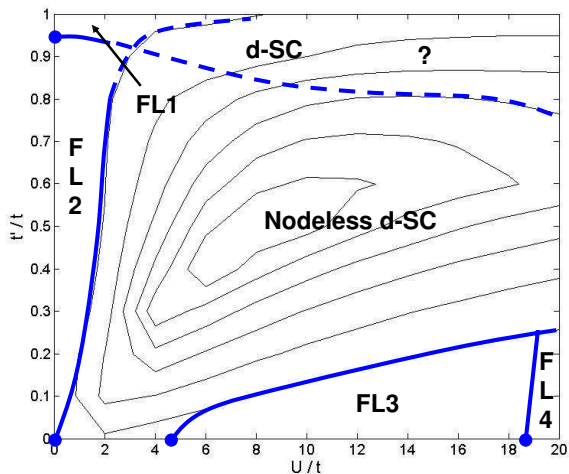} 
\caption{{\footnotesize (Color online) A speculative phase diagram of the checkerboard model at $x=1/16$.}}
\label{fig:PD_cb}
\end{figure}

 As mentioned before, computing all the excited states that enter this sum is not feasible, but a lower  bound estimate can be readily obtained in the ``single mode approximation'', by approximating the sum by the single term involving the ground state with $N-M+2$ electrons.  We refer to this as $\chi_a^{(SMA)}$.  In terms of the pair-field expectation values discussed in the previous section, 
 \ba
&& \chi_s^{(SMA)}=|\Delta_s|^2/\epsilon \\
 &&\chi_w^{(SMA)}=|\Delta_w|^2/\epsilon,\ {\rm and} \nonumber \\
 &&\chi_T^{(SMA)}=|\Delta_s + \Delta_w|^2/2\epsilon. \nonumber
 \label{SMA}
 \ea
where $\epsilon\equiv [E_0(M) - E_0(M+2) + 2\mu].$
Thus, the quantities plotted in Fig.~\ref{fig:cc_sq} can be viewed (up to a factor of $1/\epsilon$) as the SMA to the susceptibility at constant $\epsilon$.
Moreover, it is worth mentioning that they are almost exact to the lower bound susceptibility as long as $\epsilon\rightarrow 0$ by tuning chemical potential properly.
 




\section{Interpretation}

We now discuss the implications of our results.  In particular, we are interested in using the present information to infer, as much as possible, the phase diagram of the checkerboard and striped Hubbard models in the thermodynamic limit, $N\to \infty$.  The results of our analysis and some additional speculations lead us to propose the qualitative phase diagram shown in Fig.~\ref{fig:PD_cb}.

\subsection{The Checkerboard Model}
There are two limits in which the checkerboard model simplifies:  $U \ll t^\prime$, where it can be studied using 
conventional diagramatic methods, 
and $t^\prime \ll t$, where it reduces to weakly coupled squares which can be treated\cite{Tsai06,myriad} using degenerate perturbation theory in $t^\prime/t$.  This allows us to deduce the solid portions shown in the phase diagram in Fig.~\ref{fig:PD_cb} without resorting to the present numerical results.  

\begin{figure}[t]
\includegraphics[width=7.8cm]{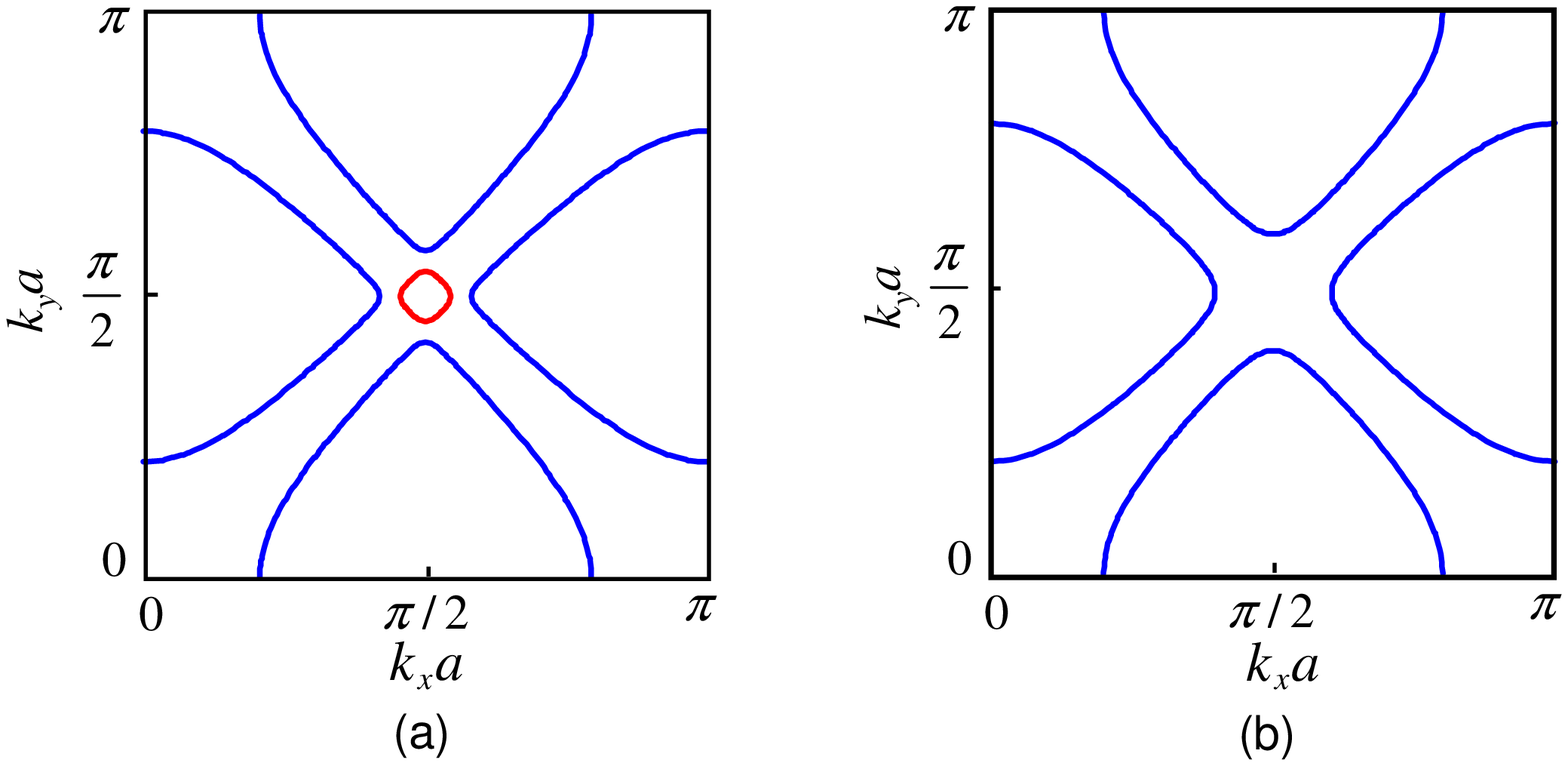} 
\caption{{\footnotesize (Color online) A sketch of the Fermi surface with $x=3/16$ for (a) $t^\prime = 0.95t$, showing the two electron and one hole Fermi pockets,
and (b)  
$t^\prime = 0.70t$, with only the two electron pockets. }}
\label{fig:Fermisurface}
\end{figure}

For $U=0$, and in the thermodynamic limit, the non-interacting Fermi-surface depends qualitatively on $t^\prime/t$. There are four bands since each unit cell contains four sites 
and $(2\pi/2a,0)$ and $(0,2\pi/2a)$ are the basis vectors of the reciprocal lattice.  
 For $t \geq t^\prime \geq t^\prime_c(x)$, there are two 
 electron pockets enclosing the $M$ points $(\pi/2a,0)$ and $(0,\pi/2a)$ respectively, and one hole pocket, enclosing the ``nodal point'' $(\pi/2a,\pi/2a)$,  
 as shown in Fig.~\ref{fig:Fermisurface}(a).  The hole pocket shrinks to a point as $t^\prime$ approaches the critical value for a Lifshitz transition, $t^\prime_c(x)$. For $0 < t' < t^\prime_c(x)$, only the two electron pockets enclosing the $M$ points remain, as shown in Fig.~\ref{fig:Fermisurface}(b). 
For $x=1/16,1/8,3/16$, and 1/4, $t_{c}^{\prime}$ is, respectively, $0.95t,0.89t,0.82t$, and $0.75t$.

This Lifshitz transition appears in the conjectural phase diagram in Fig.~\ref{fig:PD_cb} as the boundary between two Fermi liquid phases - FL1 (with two electron pockets plus one hole pocket) and FL2 (with only two electron pockets). In sketching this figure, we have assumed that the Fermi liquid phases are stable in the presence of a small repulsive $U$;  it is likely that this is not strictly the case, since there is probably a Kohn-Luttinger instability of any Fermi liquid~\cite{Kohn65}, but if this occurs, it is on such a low energy and temperature scale that it can be neglected for present purposes.


The small $t^\prime$ portion of the phase diagram in Fig.~\ref{fig:PD_cb} was derived previously in Refs.~\onlinecite{Tsai06} and \onlinecite{myriad}.  For $0 < U < U_c\approx 4.6t$, where an isolated Hubbard square has a positive pair-binding energy, there exists a nodeless d$_{(x^2-y^2)}$-wave superconducting phase. For $U_c < U < U_T\approx 18.6t$, there is a third Fermi liquid phase, FL3, which has the same Fermi surface topology as FL2.  Finally, for $U_T < U$ (where the isolated square with one doped hole has a fully polarized spin 3/2 ground state), the system exhibits an exotic, spin 3/2 Fermi liquid phase, FL4, if $x<1/8$, while it phase separates into two insulating antiferromagnetic charge ordered phases if $1/8 < x < 1/5$.     

The superconducting state at small $t^\prime$ arises from the geometry of the square and the strong correlations produced by $U$.  However, across the phase boundary between the superconducting phase and FL2 or FL3, it is reasonable to view the onset of superconductivity as a BCS-like Fermi surface instability.  In this limit, the fact that the d-wave superconductor is nodeless can be understood as being a simple consequence of the fact that the Fermi surface does not intersect the line of d-wave gap nodes, so there are no gapless quasiparticles.

To obtain a more complete phase diagram, we have used the insights obtained from the present exact diagonalization studies.  It is, of course, not clear to what extent the results from small system studies can be extrapolated to the $N\to \infty$ limit.  However, at least in the case of the pair-binding energy, since the results appear to be relatively insensitive to boundary conditions for $t^\prime < 0.8t$, we feel that we can use these results as the basis of a set of  plausible conjectures.  These are shown as the dashed lines in Fig.~\ref{fig:PD_cb}.  Notice that we have superposed the conjectured phase diagram on the contour plot of the pair-binding energy from Fig.~\ref{fig:contour_pbe}(a).  Where the pair-binding energy on the 4 $\times$ 4 system is large and insensitive to boundary conditions, we feel that we are on sound grounds when we speculate that this corresponds to a well developed, nodeless d-wave superconducting state in the $N\to\infty$ limit. 

Unfortunately, where $\Delta_{pb}$ is small and/or sensitive to boundary conditions, this could mean that in the $N\to \infty$ limit, the system has entered a gapless phase, {\it i.e.} a FL or a nodal d-wave superconductor.  However, it could simply mean that the minimum gap of a nodeless d-wave superconductor is small and the corresponding correlation length is long.  In drawing our conjectural phase diagram, we have assumed the former, and so,
to the extent possible, we have drawn phase boundaries between the nodeless d-wave phase and various 
gapless phases along contours separating the region of ``large'' and boundary condition insensitive pair-binding energy, to regions with smaller, and/or strongly boundary condition dependent pair-binding.  Clearly, the upper portion of the phase diagram ($t^\prime > 0.8t$) is the most speculative portion, including the entire region in which nodal d-wave superconductivity occurs.

The existence of a tetracritical point in the conjectured phase diagram, with the additional implication that there exists a nodal d-wave superconducting state for large enough $t^\prime$, follows from the nature of the known phases  along the edges of the phase diagram.  In particular, if there is a direct, continuous transition from FL1 to a d-wave superconductor, the d-wave superconductor must be nodal.  However, we cannot rule out the possibility of a direct first order transition from FL1 to a d-wave superconductor, in which case the tetracritical point could be replaced by a bicritical point, and the d-wave superconductor could always be nodeless.  
The portion of the phase diagram with $t^\prime>0.8t $ could also exhibit additional broken symmetry phases~\cite{wellein05,Arrigoni04,Assa02}.

\subsection{The Striped Model}

It is impossible, with any degree of confidence, to use the present results to infer anything new about the phase diagram of the striped Hubbard model in the $N\to \infty$ limit. 
For $t^{\prime}\ll t$, the model becomes an array of 
 weakly connected two-leg ladders, a problem which was studied previously by E. Arrigoni {\it et al.} in Ref.~\onlinecite{Arrigoni04}. From that work, we know that for $x<x_c \approx 0.1$, there exists a 
 nodeless ``d-wave like'' superconducting state 
over a very broad range of $U/t$.  In contrast, the oscillatory behavior of the pair-binding energy seen in the Fig.~\ref{fig:contour_pbe}(c) as $t^{\prime}\rightarrow 0$ is a special feature of the a $4\times 2$ ladder, which 
 does not 
 occur in a system of weakly coupled, infinitely-long two-leg ladders.
 Combining this observation with the strong boundary condition dependence of the pair-binding energy apparent in Fig.~\ref{fig:pbenergy}(b) (even when the twist in the boundary conditions is applied in the direction perpendicular to the stripe direction), we are forced to conclude that 
 the results on the 4 $\times$ 4 stripe lattice are not representative of the $N\to \infty$ limit.
 
We therefore do not venture to draw even a conjectural phase diagram for this system.
Nevertheless, on the basis of the fact that, in Fig.~\ref{fig:contour_pbe}(c), there is an extended region with relatively large pair-binding energy 
 when $t^{\prime}$ is a substantial fraction of $t$, 
 makes plausible the speculation made in Ref.~\onlinecite{Arrigoni04} that the nodeless superconducting state grows in strength for a substantial range of non-infinitessimal $t^\prime/t$. Furthermore, if indeed there is a nodal d-wave state for $t^\prime/t$ near 1, there must also exist a Lifshitz-type phase transition to a superconducting state with gapless quasi-particles 
 at a $U/t$ dependent critical value of the inter-ladder coupling \cite{Granath01}. 

\section{Discussion}
One issue that is often ignored in discussions of HTC is the role of the longer-range Coulomb interactions.  d-wave pairing avoids the obvious deliterious effects of the on-site Hubbard repulsion between electrons, but is known\cite{whiteandme} to be fairly sensitive to longer range repulsive interactions.  To address this issue, we have computed the effect on the pair-binding energy of a nearest-neighbor repulsion, $V$, between electrons.  Indeed, we always find that the pair-binding energy decreases, more or less linearly, with increasing $V$.  However, where the pair-binding is strong for $V=0$, it remains positive up to rather large values of $V$.  For instance, for the checkerboard lattice under optimal conditions, $U=8t$ and $t^\prime=t/2$, $\Delta_{pb}$ is an essentially linear function of $V$ which vanishes at $V\approx 1.3t$.

In conclusion, we have studied inhomogeneous Hubbard models, primarily with checkerboard and striped patterns, on a $4\times 4$ square lattice with periodic boundary conditions by exact diagonalization. 
Although the existence of the 
HTC  in the uniform Hubbard model is still a controversial issue \cite{Aimi07}, we have produced clear evidence that, without considering other non-electronic degrees of freedom such as phonons, strong pairing of electrons can be achieved from purely repulsive interactions if certain modulations of the electronic structure are introduced. Non-monotonic dependence of the pair-binding energy and the pair-field pair-field correlators on the degree of inhomogeneity  ($t^{\prime}/t$) were found to be generic. 
This observation supports the notion that there is an optimal inhomogeneity for high temperature superconductivity~\cite{sudip,Ivar05,Aryanpour07,Carlson07,Arrigoni03,Arrigoni04}.  
Since exact diagonalizaiton studies cannot access significantly larger systems, it has not been possible to carry out finite size scaling to corroborate this conclusion.  However, we hope that the present results will stimulate further work on larger system using more efficient numerical tools such as Density Matrix Renormalization Group (DMRG)~\cite{White08} or Quantum Monte Carlo (QMC).

\begin{acknowledgments}
We thank E.~Fradkin and D.~J.~Scalapino for helpful discussions. This work was supported in part by DOE Grant No. DE-FG02-06ER46287. H.Y. is supported by a SGF at Stanford University. Computational resources were provided by SLAC at Stanford University and some of the computations were performed at the CSCS Manno, Switzerland.
\end{acknowledgments}

\end{document}